\documentclass[fleqn,usenatbib, useAMS, letters]{mnras}

\usepackage{newtxtext,newtxmath}

\usepackage{graphicx}	
\usepackage{amsmath}	
\usepackage{amssymb}	
\usepackage[british]{babel}

\usepackage[T1]{fontenc}
\usepackage{ae,aecompl}

\newcommand{\kic}{KIC~8462852}

\title{KIC 8462852: Will the Trojans return in 2021?}

\author[F.~J.~Ballesteros et al.]{
Fernando~J.~Ballesteros,$^1$\thanks{E-mail: fernando.ballesteros@uv.es}
Pablo~Arnalte-Mur,$^{1,2}$ \newauthor Alberto~Fern\'andez-Soto,$^{3,4}$
and Vicent~J.~Mart\'inez$^{1,2,4}$
\\
$^1$Observatori Astron\`omic, Universitat de Val\`encia, C/ Catedr\'atico Jos\'e Beltran 2, E46980-Paterna (Val\`encia), Spain\\
$^2$Departament d'Astronomia i Astrof\'{\i}sica, Universitat de Val\`encia, E46100-Burjassot (Val\`encia), Spain \\
$^3$Instituto de F\'isica de Cantabria (CSIC-UC), E39005-Santander, Spain\\
$^4$Unidad Asociada Observatorio Astron\'omico (IFCA-UV), E46980-Valencia, Spain
}

\date{Accepted by MNRAS on 2017 June 20. Received 2017 June 20; in original form 2017 May 23}

\pubyear{2017}

\begin{document}
\label{firstpage}
\pagerange{\pageref{firstpage}--\pageref{lastpage}}
\maketitle

\begin{abstract}
KIC 8462852 stood out among more than 100,000 stars in the  {\it Kepler} catalogue because of the strange features of its light curve: a wide, asymmetric dimming taking up to 15 per cent of the light at D793 and a period of multiple, narrow dimmings happening approximately 700 days later. Several models have been proposed to account for this abnormal behaviour, most of which require either unlikely causes or a finely-tuned timing. We aim at offering a relatively natural solution, invoking only phenomena that have been previously observed, although perhaps in larger or more massive versions. We model the system using a large, ringed body whose transit produces the first dimming and a swarm of Trojan objects sharing its orbit that causes the second period of multiple dimmings. The resulting orbital period is $T\approx12$ years, with a semi-major axis $a\approx6$ au. Our model allows us to make two straightforward predictions: we expect the passage of a new swarm of Trojans in front of the star starting during the early months of 2021, and a new transit of the main object during the first half of 2023.
\end{abstract}




\section{Introduction}

In September 2015 the discovery of an extraordinary object in the {\it Kepler} field was announced: the star KIC 8462852\footnote{Also  known in the literature as {\it Boyajian's star, Tabby's star} or {\it the WTF star}.}.
Volunteer planet hunters \citep{planethunters} have expressed that its light curve is probably the most bizarre among the more than 100,000 light curves in the {\it Kepler} field. 
It is unique because it presents brief, deep drops in flux, with non-periodic repetitions and asymmetric dips, and one particularly complex event around day D1500 that covers up to 20 per cent of the stellar flux\footnote{All through this work we will refer to dates using the {\it Kepler} mission dating system, \textit{i.e.} BJD-2454833.}.
\citet{boyajian} analyzed and discussed possible astrophysical scenarios that could account for it. They discarded instrumental problems or intrinsic variability of the star or its M star dwarf supposed companion as possible causes. The authors considered different scenarios including dust from collisions within objects in a possible asteroid belt, debris from a giant collision similar to the one that supposedly caused the creation of the Earth's Moon, and a swarm of comet fragments orbiting around the star, which could account for the dips in the light curve. The latter possibility is the one that Boyajian et al. consider most likely, given that the other features are not expected in a main sequence star such as KIC~8462852. Subsequently other works by \cite{schaefer} and \cite{montet} added separate intriguing secular flux dimmings with timescales of $\sim$100 and 4 years, respectively.
Note, however, that the century-scale dimming has been disputed by some authors \citep{hippke1,hippke2}.

In this work we introduce an alternative scenario where stable debris would be expected: the Trojan regions around a giant, ringed object orbiting KIC 8462852. Most of the scenarios that have already been discussed by other authors invoke the presence of astronomical objects that range from uncommon to never directly observed, from the relatively mundane comet clouds in \citet{boyajian} to the alien Dyson sphere in \citet{wright}. Our model requires the presence of relatively familiar objects, namely a large planet with orbiting rings and a cloud of Trojan asteroids. Moreover, our model allows us to make a definite prediction: the leading Trojan cloud should induce a new period of irregularities in the light curve approximately in 2021.

All {\it Kepler} data for KIC8462852 analysed in this work were obtained from the `NASA Exoplanet Archive'\footnote{http://exoplanetarchive.ipac.caltech.edu/} web service \citep{basri}. This Letter is organized as follows: Section 2 discusses the Trojan hypothesis, we make a brief discussion of the implication of the results in Section 3, and give our conclusions in Section 4.


\section{Trojan hypothesis}
\label{Trojan}

We present our interpretation of the features observed in the light curve of KIC 8462852 as due to the transit of a large orbiting body and its Trojan cohort. The detailed properties of the body that produced the D793 event, not critical to the Trojan hypothesis we introduce in this manuscript, will be presented elsewhere (Ballesteros et al., in prep.). We mention here that the observed shape of the first transit can be reproduced by means of a large body with an extensive ring system, transiting the star at a relatively large impact parameter. The ring system is slightly tilted with respect to the orbital plane, a fact that combined with its impact parameter produces the observed temporal asymmetry.

One of our preferred hypothesis among those considered in \citet{boyajian} to explain the peculiarities in the light curve of \kic\ is that of a giant impact. According to it the second event at D1500 would be produced by the same material observed at D793, dispersed by a large impact and seen at a time corresponding to the following orbit (from where an orbital period $T \sim 700$ days would be deduced). Nevertheless, Boyajian and collaborators objected to this model based on the low probability of witnessing such an event and the non-repetition of the dips that appeared early in the Kepler mission coverage, and they preferred the hypothesis of a group of objects in a highly eccentric orbit around the star. We offer here an alternative scenario which does not require  fine-tuned time dependence, so our witnessing it does not render it improbable---it is a recurrent event. It may represent an extreme scenario, but this is to be expected given the \textit{a priori} fact that we are trying to explain a light curve selected because of its extreme rarity. 

We interpret the cluttered behaviour of the second epoch, observed from $\sim$D1500, as caused by a cohort of objects close to the L5 Lagrangian point associated to the main orbiting body. Such a stable and large swarm of bodies, debris and dust, gravitationally confined around the L5 region, can explain several of the observed features. We remark that the presence of relatively large amounts of dust is not excluded; if dust is farther away than 0.2-0.3 au from the star it wouldn't be readily detectable by WISE or {\it Spitzer} \citep[see][]{boyajian,marengo}. 

In our Solar System several worlds have gathered bodies at their Trojan regions. It is the case of the Earth, Mars, Neptune and especially Jupiter, but also of moons like Tethys and Dione; all of them sharing their orbits with objects in their Trojan points. Hydrodynamic simulations of protoplanetary disks \citep{laughlin} show that dust and disk material linger in the Trojan stable regions of a planet, remaining there after the planetary formation process is complete. Orbital instabilities such as proposed by the Nice model \citep{nice,nice2,morbidelli} could cause large numbers of small bodies to be in unusually excited orbits, which could create both Trojans and rings. Additionally, violent events in the past could also have led to the capture of objects in these regions \citep{morbidelli}. Thus, stable Trojan bodies related to exoplanets should be commonplace. In fact, \citet{Hippke2015a} already found a hint of the presence of Trojans in long-period exoplanetary orbits using {\it Kepler} data.

In order to estimate the extension of the Trojan swarm along the orbit we can examine the case of Jupiter. \cite{jewitt} measured an apparent FWHM of the L4 swarm along the ecliptic of 26.4$^\circ$ $\pm$ 2.1$^\circ$, corresponding to a linear size of 2.4 au. Taking twice this value to encompass the whole swarm, we can estimate that Jupiter's L4 Trojan cloud roughly covers an angular extent of $\sim$50$^\circ$ along its orbit, which is also in accordance with \cite{karlsson}. Regarding the strange structure of the second transit epoch, we remark that studies of stability for planets more massive than Jupiter with elliptic orbits \citep{erdi} show that the spatial distribution of the stability regions around L4 and L5 could have a very complex structure. 

In the Solar System we observe that the ensemble mass of the Jupiter Trojans is $\sim 0.0001 M_{\earth}$, with a total cross section similar to a disk of radius $\sim 2000$ km (i.e. larger than that of the Moon), but a mass which is two orders of magnitude smaller \citep{jewitt, sheppard}.
Even though in our Solar System the Trojan-to-planet mass ratio is low ($\le 10^{-8}$), extrasolar planets may have more massive Trojans, perhaps even with ratios as large as one-to-one \citep{ford}. 

Taking as an example a hypothetical amount of mass $M = M_{\rm Jup}$ trapped in the Trojan regions of KIC 8462852 following the same size distribution as the Jupiter Trojans \citep{wong,Fernandez2003,Harris1997}, we reach an effective cross section as high as $\sim 200$ times the star cross section when considering bodies with diameter larger than 1 km. Allowing for Trojan bodies as small as 50 meters the figure can rise up to 500 times the cross section of the star. Nevertheless, large amounts of dust can remarkably relax any mass requirement: dust accumulated in the Trojan region can be a major factor for opacity, in fact dominating the cross section. \citet{boyajian} estimate the mass of dust necessary to produce the observed opacity to be $6.7 \times 10^{18}$g. As we will discuss in the next Section, using this same model we obtain that a cloud mass well below 10$^{-4}$ M$_{\earth}$ within the Trojan regions would be enough to produce the observed results.  Collisions among Trojans or catastrophic events in the past could have generated or trapped large enough amounts of dust in these regions.

One problem remains: the largest individual dip observed close to D1500, which shows substructure in the light curve, would correspond to a large event which covered a significant fraction of the stellar cross section. The presence of substructure points to the possibility that we may be observing a clustering of smaller bodies and dust, probably gravitationally linked. This is indeed one of the most difficult aspects that any model of this fascinating observation ought to address, but, as in Boyajian's cometary scenario, clumps seem natural for the Trojan model and the existence of collisional families (that can produce an additional amount of dust, thus increasing the cross section) should be expected.
In our case we need to postulate the presence of one such cluster close to the Lagrange L5 point of the main planet. As mentioned above, some studies point that this may not be particularly strange \citep{ford}. 

\begin{figure*}
   \centering
   \includegraphics[width=0.5\hsize]{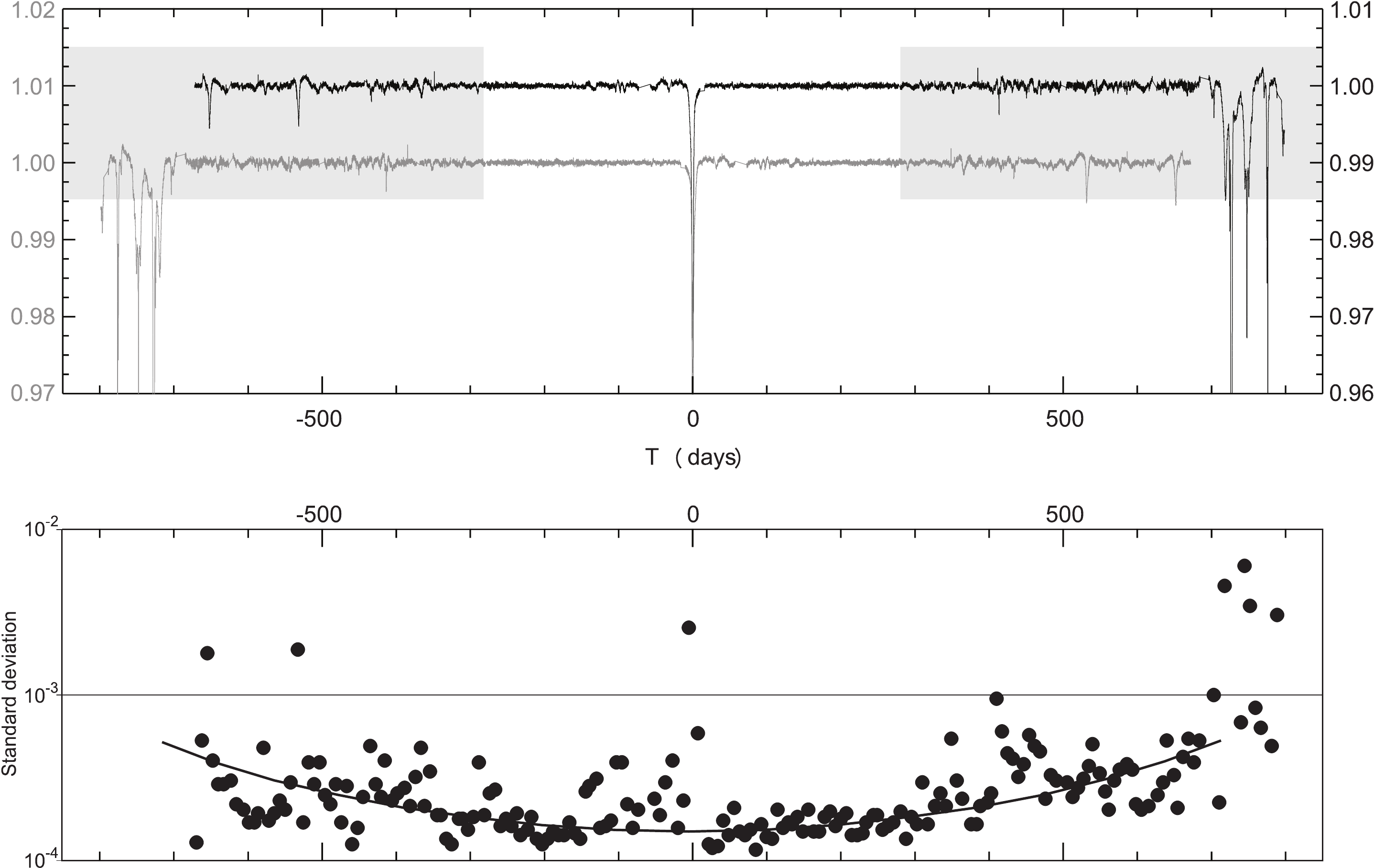}
      \caption{Top: Light curve of KIC 8462852, inverted around D793 (gray) and superimposed on the original (black). The time-inverted curve is shifted downwards for clarity. There is a quiescent period lasting $\approx 600$ days centered around D793. Outside this period we detect an increase in the variability of the light curve (shaded regions). Bottom: Standard deviation of the data. Each point represents the standard deviation of the light intensities during a one-week period, and seems to increase symmetrically as one moves away from D793. The solid line is a guide to the eye and does not represent a fit to the data. Points with large deviations correspond to light curve peaks.}
         \label{signature}
   \end{figure*}

All in all, according to the hypothesis that the features observed at $\sim$D1500 correspond to the passage of Trojans close to the trailing L5 point of the transiting planet, one should expect a similar signature in the symmetric point before the planetary transit.
Unfortunately {\it Kepler} observations of KIC 8462852 started at D120, just after the epoch when the putative symmetric L4 Trojans would have transited (ending around D65).
Notice, however, that at the beginning of the time series there are some small features that could be produced by the last Trojans inhabiting the L4 zone.
In fact, reversing the time series around D793 (see Figure \ref{signature}, top), a certain symmetry with the original time series appears.
Beyond a distance of $\sim 300$ days before and after this date the variability of the time series seems to increase, while it remains quieter in the region in between. 
Calculation of the time series standard deviation in weekly periods seems to confirm this symmetry around the presumed planet, strengthening the Trojan hypothesis.
We have confirmed that during the same period other stars in the nearby region of the CCD where KIC~8462852 was detected did not show this kind of fluctuations, ruling out the hypothesis of variations in the CCD sensitivity with time, changes in the illumination on the focal plane, or other instrumental effects that should have affected all targets approximately in the same manner.
One thing we cannot exclude, however, is that some of the minor dips that we are assuming could be caused by either trailing or leading Trojans could in reality be caused by dimming episodes in nearby, blended objects, as discussed by \citet{2016ApJ...833...78M}. We remark, though, that because of their aperiodic nature we would be somehow transferring the problem of explaining their nature from KIC 8462852 to the other objects.

\section{Implications of the model}

If the Trojan hypothesis is correct we can obtain a direct estimate of the orbital period. The time interval between the first and second main events in the {\it Kepler} data is approximately two years. Assuming an orbital separation of $\sim 60 ^\circ$ between them, the orbital period would be $\sim 12$ years, which given the mass estimated for the star \cite[1.43 $M_{\odot}$, ][]{boyajian} would imply a semi-major axis of 5.9 au. With these parameters and the stellar radius $R=1.58\, R_{\odot}$ estimated by \cite{boyajian}, an object in a circular orbit would move at 15 km s$^{-1}$ and an equatorial transit would last $\sim 1.7$ days. Considering a $\sim$20 per cent uncertainty in the mass and radius of the star as \cite{boyajian} quote in their paper, and the fact that an elliptical orbit would induce an extra uncertainty in the transit speed\footnote{A large orbital eccentricity would also introduce instability in the Trojan regions. Previous works about this indicate that an eccentricity as high as 0.3 \citep{chanut, robutel} or even 0.6 \citep{erdi} could be allowed for.}, the transit speed and duration could accommodate changes of up to a factor of 2. These figures move the expected duration of the main planetary transit closer to the one observed, although tensions persist: \citet{boyajian} estimate a velocity for the transiting material of up to 50 km s$^{-1}$, and a particular transiting event close to D1568 had a duration as short as 0.4 days. The above-mentioned uncertainties together with the effect of a large impact parameter and the possible libration of the Trojans around their equilibrium points could diminish the tension between both values.

If we consider that in our model the hypothetical Trojan cloud started at the epoch when the light curve of KIC 8462852 began to increase its variability (shaded region in Fig. \ref{signature}), the duration until the end of the light curve is $\sim$500 days, which would represent an angular extent of $\sim$40$^\circ$ for the assumed orbital period of $\sim$12 years. Given that the irregular behaviour very probably extended beyond the end of the observed light curve, the extension could perfectly reach a value comparable to $\sim$50$^\circ$, the orbital extension of the Trojan swarm around Jupiter's L4 point presented in the previous Section.

We have used a very simple model, a bi-Gaussian distribution of material around the L5 point with a FWHM extension of 26.4 degrees along the orbit and the same spread in the vertical direction, to estimate that only $\approx1/350$ of the cloud material will cross in front of the projected area of the star. This implies the amount of dust and material in the whole proposed Trojan region could be $\sim350$ times bigger than that producing the transit, or 700 times bigger considering both Lagrange regions. As the infrared luminosity limits from WISE and Spitzer presented in \citet[][see their Figure 12]{boyajian} only refer to dust passing directly in front of the star, the luminosity estimate at 6 au should then be increased by a factor 700. This is still well below the observational WISE limits, that could accomodate a factor up to 2000. Moreover the dust model in \citet{boyajian} rescaled to 6 au and distributed according to the former bi-Gaussian model yields an mass estimate M=$4 \times 10^{23}$g if we allow particle sizes up to 1 cm---a mass of dust well below 10$^{-4}\, M_{\earth}$ within both Trojan regions.  

\cite{boyajian} performed an estimate of the range of planetary masses and orbital periods that could be compatible with the observed absence of radial velocity variations in the available spectroscopic observations.
We remark that these are scarce (four runs), of not very high precision ($\sigma_v \sim 400 $ m s$^{-1}$), and cover a relatively small period of time (less than 1.5 years).
We have performed a similar calculation, fixing the orbital period at $T=12$ years and assuming that the D793 transit happened close to the periastron.
Under this assumptions, and allowing for orbital ellipticities between 0 and 0.6, we derive a $1\sigma$ upper limit to the mass of the planetary object $M_{\rm p} < 130-170\, M_{\rm Jup}$, with the highest limit corresponding to circular orbits.
This limit does hardly constrain the model at all.
In fact, just requiring the stability of the Trojans near the Lagrangian points imposes a ratio $M_{\rm star}/M_{\rm p} > 25$ which, given the estimated mass of the star, translates into $M_{\rm p} \lesssim 60 M_{\rm Jup}$, a more restrictive limit.
New determinations of the radial velocity, even if they were of similar precision, would greatly reduce the uncertainty which is mostly caused by the degeneracy between the systemic radial velocity and the maximum $\Delta v$ induced by the orbiting body. A stringent limit on the mass of the orbiting body would not be critical for the Trojan model here presented, as our estimates depend on apparent sizes derived from observed opacities, large values of which can be produced by relatively small masses of dust and/or small bodies, and a probable highly opaque ring system.

\subsection{The dimming event at D3060}

On May 19, 2017 an alert was launched by T. Boyajian reporting the observation of a possible low-intensity dip in the lightcurve of KIC 8462852 \citep{Boyajian2017,Boyajian2017a}. Observers  interpreted this dip as the possible beginning of a new passage of the debris associated to the catastrophic event that may have happened to the large planetary object that transited at D793, causing it to be observed as a swarm at D1500, confirming in such case an orbital period of approximately two years. 

Within our model the timing of this event would correspond approximately to the opposition of the main body, therefore the corresponding L3 point would be passing in front of the star by this time. 
In this scenario, this dimming event could be explained by the effect of objects akin to the Hildian asteroids in the Solar System passing the L3 area\footnote{The Hildas are a dynamic group of Solar System asteroids that occupy a 3:2 orbital resonance with Jupiter. The resonance makes their aphelia alternatively close to Jupiter's Lagrange points L3, L4, and L5; in such a way that at any time they are preferentially found close to those three points.}. If what we have identified as Trojan asteroid regions are really so densely populated, in principle one could expect also a rather high density of Hildas, particularly if some catastrophic event in the (relatively) near past has populated both orbits. In this case, we would expect a few dimming events concentrated around the time when L3 transits in front of the star, as well as the possibility that some of the events that we have generally classified as Trojans could in reality be due to Hildas. This hypothesis  could help reconcile some of the fastest ones, as the Hilda orbits have a shorter period. The concentration of Hilda events close to L3 would, in any case, be much smaller than the complex event at D1500 caused by the passage of the Trojans.

\section{Conclusions}

\begin{figure*}
   \centering
   \includegraphics[width=0.6\hsize]{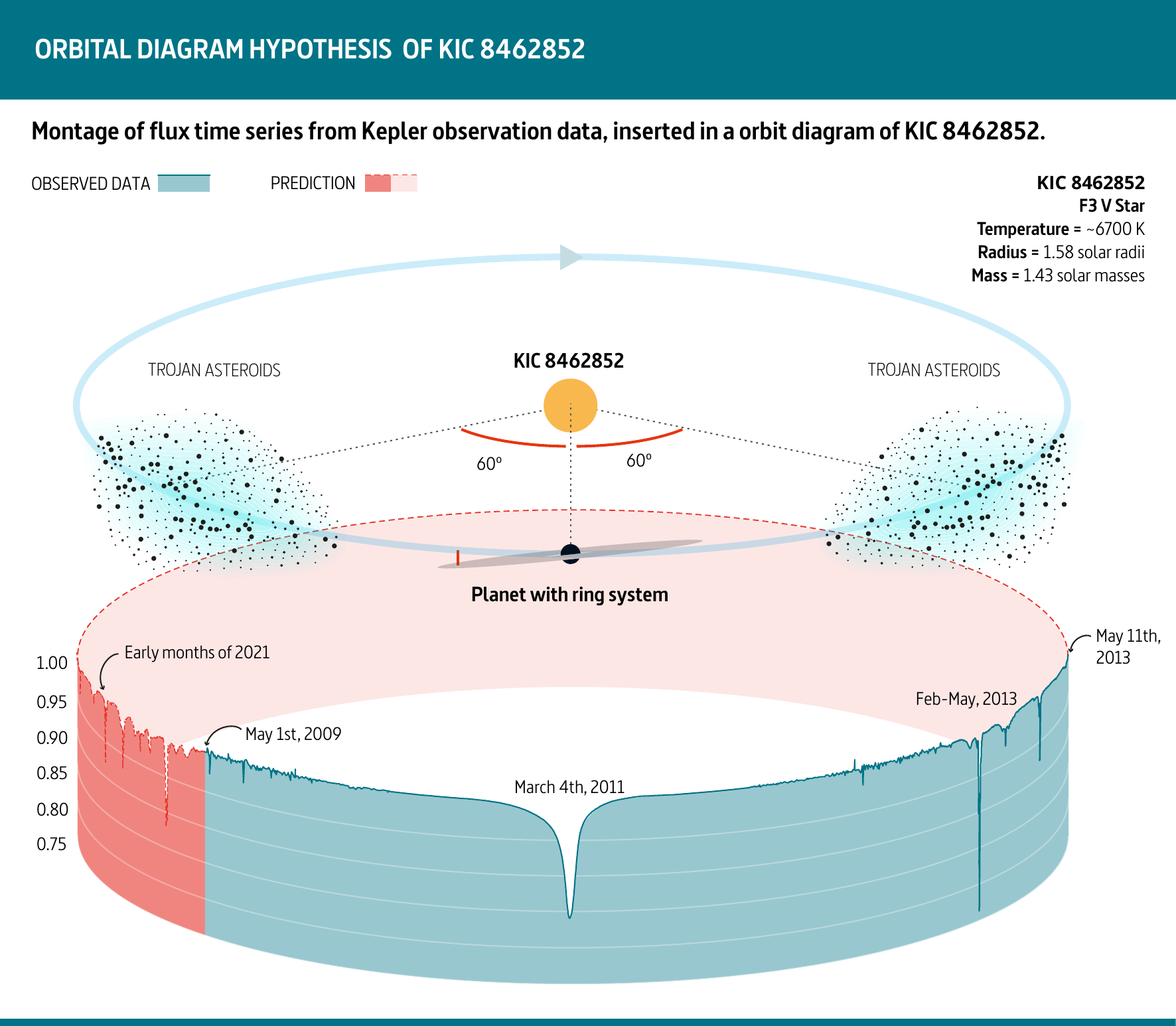}
      \caption{Diagram showing a hypothetical ringed giant body orbiting the star, together with its dense populations of Trojan bodies and dust around the L4 and L5 Lagrange points. We also present below it the observed (blue) and expected (red) light curve.}
         \label{3dmodel}
   \end{figure*}

We have presented an alternative model to explain the odd appearance of the light curve of KIC 8462852. We propose that a grazing transit of a large, ringed object could produce the asymmetric transit observed at D793, whereas a huge swarm of Trojan objects inhabiting its L5 orbital point could have caused the irregular transit at D1500. We deduce an orbital period $T\approx$ 12 years. In principle this would imply a transit speed slower than observed, although a highly eccentric orbit would make them close to compatible. Full details of the modeling of the main planetary transit at D793 will be presented elsewhere. We estimate its mass to be $\lesssim 150\, M_{\rm Jup}$ (stellar radial velocity) and $\lesssim 60\, M_{\rm Jup}$ (Trojan cloud dynamic stability), but it can be much smaller as the observed large cross section does not necessarily imply a large mass.
We show in Figure \ref{3dmodel} a diagram representing the main parameters and properties of our model, together with an idealized vision of the light curve. Note that our model does not explain the secular dimmings observed by \cite{schaefer} and \cite{montet}. Such effects would call for a completely unrelated explanation.

Given the exceptional behaviour of this light curve, our explanation is also somehow exceptional --as are all the other proposed hypotheses-- but not too unconventional. It is nourished by the evidence of similar (although obviously not identical) existing objects in our Solar System and beyond, which have been previously studied in detail. A key advantage of our model is the necessary repeatability of the phenomenon. This fact puts aside any coincidence or temporal fine-tuning quandary, as our observation does not imply a particular moment in the history of the observed system. 

This repeatability also allows us to carry out a testable prediction: under the assumptions of our hypothesis, considering an orbital period of $\sim 12$ years, and taking into account that the region of deepest dimming lasted three months, we predict the onset of a new epoch of irregular transits at $\sim$D4430, i.e., February 2021. In other words, during the early months of 2021 the swarm of objects at the symmetric L4 Lagrangian point will transit the star, starting an epoch of cluttered dimmings. Two years later, during the first half of 2023, we expect a new transit of the ringed planetary body.

\section*{Acknowledgements}
We thank D.\ Fabrycky, D.\ Richards and J.\ Wright for pointing out an error regarding the possibility of a secondary eclipse in a previous version of our work, J.\ Hopkins for indicating the relevance of Hildas in this scenario and D.\ Ragozzine and D.\ Angerhausen for useful comments. VJM thanks the estimulating panel discussion on a preliminary draft of this manuscript held in April 2017 at the Geological Museum (University of Copenhagen) with
the Astrophysics and Planetary Science group led by U.G.\ Jorgensen. We gratefully thank our anonymous referee for his/her insightful comments that have improved our manuscript.
This work was supported by Spanish MINECO projects AYA2013-48623-C2-2 and AYA2016-81065-C2-2, including FEDER contributions and by the Generalitat Valenciana project of excellence PROMETEOII/2014/060. This paper includes data collected by the {\it Kepler} mission. Funding for the {\it Kepler} mission is provided by the NASA Science Mission directorate.




\bibliographystyle{mnras}
\bibliography{ref}

\begin{thebibliography}{}
\makeatletter
\relax
\def\mn@urlcharsother{\let\do\@makeother \do\$\do\&\do\#\do\^\do\_\do\%\do\~}
\def\mn@doi{\begingroup\mn@urlcharsother \@ifnextchar [ {\mn@doi@}
  {\mn@doi@[]}}
\def\mn@doi@[#1]#2{\def\@tempa{#1}\ifx\@tempa\@empty \href
  {http://dx.doi.org/#2} {doi:#2}\else \href {http://dx.doi.org/#2} {#1}\fi
  \endgroup}
\def\mn@eprint#1#2{\mn@eprint@#1:#2::\@nil}
\def\mn@eprint@arXiv#1{\href {http://arxiv.org/abs/#1} {{\tt arXiv:#1}}}
\def\mn@eprint@dblp#1{\href {http://dblp.uni-trier.de/rec/bibtex/#1.xml}
  {dblp:#1}}
\def\mn@eprint@#1:#2:#3:#4\@nil{\def\@tempa {#1}\def\@tempb {#2}\def\@tempc
  {#3}\ifx \@tempc \@empty \let \@tempc \@tempb \let \@tempb \@tempa \fi \ifx
  \@tempb \@empty \def\@tempb {arXiv}\fi \@ifundefined
  {mn@eprint@\@tempb}{\@tempb:\@tempc}{\expandafter \expandafter \csname
  mn@eprint@\@tempb\endcsname \expandafter{\@tempc}}}

\bibitem[\protect\citeauthoryear{{Basri}, {Borucki}  \& {Koch}}{{Basri}
  et~al.}{2005}]{basri}
{Basri} G.,  {Borucki} W.~J.,   {Koch} D.,  2005, \mn@doi [\nar]
  {10.1016/j.newar.2005.08.026}, \href
  {http://adsabs.harvard.edu/abs/2005NewAR..49..478B} {49, 478}

\bibitem[\protect\citeauthoryear{{Boyajian}}{{Boyajian}}{2017}]{Boyajian2017}
{Boyajian} T.~{\tt (@tsboyajian)}.,  2017, "\#TabbysStar IS DIPPING!
  OBSERVE!!". 19 May 2017, 4:32 AM. Tweet

\bibitem[\protect\citeauthoryear{{Boyajian} et~al.,}{{Boyajian}
  et~al.}{2016}]{boyajian}
{Boyajian} T.~S.,  et~al., 2016, \mn@doi [\mnras] {10.1093/mnras/stw218}, \href
  {http://adsabs.harvard.edu/abs/2016MNRAS.457.3988B} {457, 3988}

\bibitem[\protect\citeauthoryear{Boyajian et~al.,}{Boyajian
  et~al.}{2017}]{Boyajian2017a}
Boyajian T.,  et~al., 2017, The Astronomer's Telegram, 10405

\bibitem[\protect\citeauthoryear{{Chanut}, {Tsuchida}  \& {Winter}}{{Chanut}
  et~al.}{2004}]{chanut}
{Chanut} T.~G.~G.,  {Tsuchida} M.,   {Winter} O.~C.,  2004, in Advances in
  space dynamics 4: Celestial mechanics and astronautics. pp 57--63

\bibitem[\protect\citeauthoryear{{Erdi}, {Frohlich}, {Nagy}  \& Zs.}{{Erdi}
  et~al.}{2007}]{erdi}
{Erdi} B.,  {Frohlich} G.,  {Nagy} I.,   Zs. S.,  2007, in Proc. of the 4th
  Austrian Hungarian Workshop on celestial mechanics. p.~85,
  \mn@doi{10.1111/j.1749-6632.1980.tb15927.x}

\bibitem[\protect\citeauthoryear{{Fern{\'a}ndez}, {Sheppard}  \&
  {Jewitt}}{{Fern{\'a}ndez} et~al.}{2003}]{Fernandez2003}
{Fern{\'a}ndez} Y.~R.,  {Sheppard} S.~S.,   {Jewitt} D.~C.,  2003, \mn@doi
  [\aj] {10.1086/377015}, \href
  {http://adsabs.harvard.edu/abs/2003AJ....126.1563F} {126, 1563}

\bibitem[\protect\citeauthoryear{{Fischer} et~al.,}{{Fischer}
  et~al.}{2012}]{planethunters}
{Fischer} D.~A.,  et~al., 2012, \mn@doi [\mnras]
  {10.1111/j.1365-2966.2011.19932.x}, \href
  {http://adsabs.harvard.edu/abs/2012MNRAS.419.2900F} {419, 2900}

\bibitem[\protect\citeauthoryear{{Ford} \& {Gaudi}}{{Ford} \&
  {Gaudi}}{2006}]{ford}
{Ford} E.~B.,  {Gaudi} B.~S.,  2006, \mn@doi [\apjl] {10.1086/510235}, \href
  {http://adsabs.harvard.edu/abs/2006ApJ...652L.137F} {652, L137}

\bibitem[\protect\citeauthoryear{{Gomes}, {Levison}, {Tsiganis}  \&
  {Morbidelli}}{{Gomes} et~al.}{2005}]{nice}
{Gomes} R.,  {Levison} H.~F.,  {Tsiganis} K.,   {Morbidelli} A.,  2005, \mn@doi
  [\nat] {10.1038/nature03676}, \href
  {http://adsabs.harvard.edu/abs/2005Natur.435..466G} {435, 466}

\bibitem[\protect\citeauthoryear{{Harris} \& {Harris}}{{Harris} \&
  {Harris}}{1997}]{Harris1997}
{Harris} A.~W.,  {Harris} A.~W.,  1997, \mn@doi [\icarus]
  {10.1006/icar.1996.5664}, \href
  {http://adsabs.harvard.edu/abs/1997Icar..126..450H} {126, 450}

\bibitem[\protect\citeauthoryear{{Hippke} \& {Angerhausen}}{{Hippke} \&
  {Angerhausen}}{2015}]{Hippke2015a}
{Hippke} M.,  {Angerhausen} D.,  2015, \mn@doi [\apj]
  {10.1088/0004-637X/811/1/1}, \href
  {http://adsabs.harvard.edu/abs/2015ApJ...811....1H} {811, 1}

\bibitem[\protect\citeauthoryear{{Hippke}, {Angerhausen}, {Lund}, {Pepper}  \&
  {Stassun}}{{Hippke} et~al.}{2016}]{hippke1}
{Hippke} M.,  {Angerhausen} D.,  {Lund} M.~B.,  {Pepper} J.,   {Stassun} K.~G.,
   2016, \mn@doi [\apj] {10.3847/0004-637X/825/1/73}, \href
  {http://adsabs.harvard.edu/abs/2016ApJ...825...73H} {825, 73}

\bibitem[\protect\citeauthoryear{{Hippke} et~al.,}{{Hippke}
  et~al.}{2017}]{hippke2}
{Hippke} M.,  et~al., 2017, \mn@doi [\apj] {10.3847/1538-4357/aa615d}, \href
  {http://adsabs.harvard.edu/abs/2017ApJ...837...85H} {837, 85}

\bibitem[\protect\citeauthoryear{{Jewitt}, {Trujillo}  \& {Luu}}{{Jewitt}
  et~al.}{2000}]{jewitt}
{Jewitt} D.~C.,  {Trujillo} C.~A.,   {Luu} J.~X.,  2000, \mn@doi [\aj]
  {10.1086/301453}, \href {http://adsabs.harvard.edu/abs/2000AJ....120.1140J}
  {120, 1140}

\bibitem[\protect\citeauthoryear{{Karlsson}}{{Karlsson}}{2010}]{karlsson}
{Karlsson} O.,  2010, \mn@doi [\aap] {10.1051/0004-6361/200912629}, \href
  {http://adsabs.harvard.edu/abs/2010A%26A...516A..22K} {516, A22}

\bibitem[\protect\citeauthoryear{{Laughlin} \& {Chambers}}{{Laughlin} \&
  {Chambers}}{2002}]{laughlin}
{Laughlin} G.,  {Chambers} J.~E.,  2002, \mn@doi [\aj] {10.1086/341173}, \href
  {http://adsabs.harvard.edu/abs/2002AJ....124..592L} {124, 592}

\bibitem[\protect\citeauthoryear{{Makarov} \& {Goldin}}{{Makarov} \&
  {Goldin}}{2016}]{2016ApJ...833...78M}
{Makarov} V.~V.,  {Goldin} A.,  2016, \mn@doi [\apj]
  {10.3847/1538-4357/833/1/78}, \href
  {http://adsabs.harvard.edu/abs/2016ApJ...833...78M} {833, 78}

\bibitem[\protect\citeauthoryear{{Marengo}, {Hulsebus}  \& {Willis}}{{Marengo}
  et~al.}{2015}]{marengo}
{Marengo} M.,  {Hulsebus} A.,   {Willis} S.,  2015, \mn@doi [\apjl]
  {10.1088/2041-8205/814/1/L15}, \href
  {http://adsabs.harvard.edu/abs/2015ApJ...814L..15M} {814, L15}

\bibitem[\protect\citeauthoryear{{Montet} \& {Simon}}{{Montet} \&
  {Simon}}{2016}]{montet}
{Montet} B.~T.,  {Simon} J.~D.,  2016, \mn@doi [\apjl]
  {10.3847/2041-8205/830/2/L39}, \href
  {http://adsabs.harvard.edu/abs/2016ApJ...830L..39M} {830, L39}

\bibitem[\protect\citeauthoryear{{Morbidelli}, {Levison}, {Tsiganis}  \&
  {Gomes}}{{Morbidelli} et~al.}{2005}]{morbidelli}
{Morbidelli} A.,  {Levison} H.~F.,  {Tsiganis} K.,   {Gomes} R.,  2005, \mn@doi
  [\nat] {10.1038/nature03540}, \href
  {http://adsabs.harvard.edu/abs/2005Natur.435..462M} {435, 462}

\bibitem[\protect\citeauthoryear{{Robutel} \& {Gabern}}{{Robutel} \&
  {Gabern}}{2006}]{robutel}
{Robutel} P.,  {Gabern} F.,  2006, \mn@doi [\mnras]
  {10.1111/j.1365-2966.2006.11008.x}, \href
  {http://adsabs.harvard.edu/abs/2006MNRAS.372.1463R} {372, 1463}

\bibitem[\protect\citeauthoryear{{Schaefer}}{{Schaefer}}{2016}]{schaefer}
{Schaefer} B.~E.,  2016, \mn@doi [\apjl] {10.3847/2041-8205/822/2/L34}, \href
  {http://adsabs.harvard.edu/abs/2016ApJ...822L..34S} {822, L34}

\bibitem[\protect\citeauthoryear{{Sheppard} \& {Trujillo}}{{Sheppard} \&
  {Trujillo}}{2010}]{sheppard}
{Sheppard} S.~S.,  {Trujillo} C.~A.,  2010, \mn@doi [\apjl]
  {10.1088/2041-8205/723/2/L233}, \href
  {http://adsabs.harvard.edu/abs/2010ApJ...723L.233S} {723, L233}

\bibitem[\protect\citeauthoryear{{Tsiganis}, {Gomes}, {Morbidelli}  \&
  {Levison}}{{Tsiganis} et~al.}{2005}]{nice2}
{Tsiganis} K.,  {Gomes} R.,  {Morbidelli} A.,   {Levison} H.~F.,  2005, \mn@doi
  [\nat] {10.1038/nature03539}, \href
  {http://adsabs.harvard.edu/abs/2005Natur.435..459T} {435, 459}

\bibitem[\protect\citeauthoryear{{Wong}, {Brown}  \& {Emery}}{{Wong}
  et~al.}{2014}]{wong}
{Wong} I.,  {Brown} M.~E.,   {Emery} J.~P.,  2014, \mn@doi [\aj]
  {10.1088/0004-6256/148/6/112}, \href
  {http://adsabs.harvard.edu/abs/2014AJ....148..112W} {148, 112}

\bibitem[\protect\citeauthoryear{{Wright}, {Cartier}, {Zhao}, {Jontof-Hutter}
  \& {Ford}}{{Wright} et~al.}{2016}]{wright}
{Wright} J.~T.,  {Cartier} K.~M.~S.,  {Zhao} M.,  {Jontof-Hutter} D.,   {Ford}
  E.~B.,  2016, \mn@doi [\apj] {10.3847/0004-637X/816/1/17}, \href
  {http://adsabs.harvard.edu/abs/2016ApJ...816...17W} {816, 17}

\makeatother
\end{thebibliography}







\bsp	
\label{lastpage}
\end{document}